\def\F        {{$^{19}$F \/}}

\def\CaF      {{CaF$_2$ \/}}

\def\eg       {{\it e.g. \/}}
\def\ie       {{\it i.e.}}

\newcommand{\ee}[1]{\cdot10^{#1}}
\newcommand{\mr}[1]{\mathrm{#1}}
\newcommand{\unit}[1]{\,\mathrm{#1}}
\newcommand{\um}{\,\mu{\rm m}}
\newcommand{\us}{\,\mu{\rm s}}
\newcommand{\kT}{k_{\rm B}T}

\newcommand{\tm}{\tau_\mr{m}}

\newcommand{\Snoise}{S_{\rm noise}}
\newcommand{\tr}{\tau_\mr{r}}
\newcommand{\tottime}{\tau_{\rm{total}}}
\newcommand{\SNR}{\mr{SNR}}

\newcommand{\DN}{\Delta N}
\newcommand{\avgDN}{\overline{\Delta N}}
\newcommand{\stdDN}{\sigma_{\Delta N}}
\newcommand{\varDN}{\sigma^2_{\Delta N}}
\newcommand{\ssDN}{s_{\Delta N}^2}
\newcommand{\stdssDN}{\sigma_{s^2_{\Delta N}}}

\newcommand{\varX}{\sigma^2_X}
\newcommand{\varY}{\sigma^2_Y}
\newcommand{\varspin}{\sigma^2_\mr{spin}}
\newcommand{\varnoise}{\sigma^2_\mr{noise}}
\newcommand{\varvarspin}{\sigma^4_\mr{spin}}
\newcommand{\varvarnoise}{\sigma^4_\mr{noise}}

\newcommand{\ssX}{s_{X}^2}
\newcommand{\ssY}{s_{Y}^2}
\newcommand{\ssspin}{s_{\mr{spin}}^2}

\newcommand{\varssX}{\sigma_{\ssX}^2}
\newcommand{\varssY}{\sigma_{\ssY}^2}
\newcommand{\stdssX}{\sigma_{\ssX}}
\newcommand{\stdssY}{\sigma_{\ssY}}

\newcommand{\stdssspin}{\sigma_{\ssspin}}

\documentclass[prl,aps,twocolumn,showpacs]{revtex4}

\usepackage[dvips]{graphicx}
\usepackage{dcolumn}
\usepackage{bm}
\usepackage{epsfig}
\usepackage{amssymb}
\usepackage[colorlinks=true, pdfstartview=FitV, linkcolor=blue, citecolor=blue, urlcolor=blue]{hyperref}

\begin{document}
\global\emergencystretch = .1\hsize 

\title{The role of spin noise in the detection of nanoscale ensembles of nuclear spins}

\author{C.L. Degen$^1$, M. Poggio$^{1,2}$, H.J. Mamin$^1$, and D. Rugar$^1$}
\email{rugar@almaden.ibm.com}
\affiliation{
   $^1$IBM Research Division, Almaden Research Center, 650 Harry Road, San Jose, CA 95120, USA. \\
   $^2$Center for Probing the Nanoscale, Stanford University, 476 Lomita Hall, Stanford, CA 94305, USA.}

\date{\today}

\begin{abstract}
When probing nuclear spins in materials on the nanometer scale, random fluctuations of the spin polarization will exceed the mean Boltzmann
polarization for sample volumes below about $(100\unit{nm})^3$. In this work,
we use magnetic resonance force microscopy to observe nuclear spin fluctuations in real time.
We show how reproducible measurements of the polarization variance can be obtained
by controlling the spin correlation time and rapidly sampling a large number of independent spin configurations.
A protocol to periodically randomize the spin ensemble is demonstrated,
allowing significant improvement in the signal-to-noise ratio for nanometer-scale magnetic resonance imaging.
\end{abstract}


\maketitle

Magnetic resonance signals detected in conventional magnetic resonance imaging (MRI) experiments originate from the slight alignment of nuclear spins induced by an external magnetic field.  This thermal equilibrium (``Boltzmann'') polarization gives rise to a mean fractional spin polarization that is typically quite small,
$\avgDN/N<10^{-4}$. For the large ensembles of nuclear spins detected in MRI experiments, usually $N>10^{15}$ spins, the Boltzmann polarization is the dominant source of spin alignment. However, as new techniques, such as magnetic resonance force microscopy (MRFM) \cite{sidles95,mamin07}, push detectable volumes below $(100\unit{nm})^3$, another type of polarization becomes increasingly important: the ``statistical'' polarization.

Statistical polarization arises from the incomplete cancellation of randomly oriented spins.
The instantaneous polarization, \ie ~the difference $\DN$ between spin-up and spin-down populations, can be either positive or negative and will fluctuate on a time scale that depends on the random flip rate of the spins (for example, due to spin-lattice relaxation).
For a random ensemble of spin-1/2 nuclei, it follows from the properties of the  binomial distribution
that the statistical fluctuations have variance $\varDN=N[1-(\avgDN/N)^2]$, where the overbar indicates mean value.
In the limit of small mean polarization, which is representative of most experiments, the variance simplifies to $\varDN=N$.
The existence of statistical polarization was pointed out by Bloch in his seminal paper on nuclear induction \cite{bloch46},
and has been observed experimentally by a number of techniques, including superconducting quantum interference devices \cite{sleator85}, conventional magnetic resonance detection \cite{mccoy89,gueron89,muller06}, optical techniques \cite{crooker04}, and MRFM \cite{mamin05,mamin07}. 

As detection volumes enter the nanometer-scale regime, the standard deviation of the polarization fluctuations $\stdDN$ can easily exceed the Boltzmann polarization $\avgDN=N\mu B/\kT$, where $\mu$  is the magnetic moment of the spin, $B$ is the polarizing magnetic field and $T$ is the temperature \cite{abragam61}. The dominance of statistical polarization, as defined by $\stdDN>\avgDN$, occurs for sample volumes $V<(\mu B/\kT)^{-2} \rho_N^{-1}$, where $\rho_N=N/V$ is the spin number density. Assuming conditions representative of high-field MRI microscopy of protons in water ($B=10\unit{T}$, $T=295\unit{K}$ and $\rho_N=7\ee{28}\unit{m^{-3}}$), the volume corresponds to $\sim(230\unit{nm})^3$.  For MRFM detection of \F nuclei in calcium fluoride, as considered in this paper ($B=2.9\unit{T}$, $T=4.5\unit{K}$ and $\rho_N=5\ee{28}\unit{m^{-3}}$), the volume for statistical polarization dominance is $\sim(37\unit{nm})^3$.

Given that statistical polarization is such a strong feature of nanoscale nuclear spin detection, it is worthwhile to consider efficient methods to harness it for imaging applications. In order to generate a signal that is proportional to the spin density of the sample,
it is natural to consider using the variance as the ``signal''.
If the polarization $\DN$ is measured for $n$ independent configurations of the spin ensemble, the sample variance $\ssDN$ is estimated as
\begin{equation}
\ssDN=\frac{1}{n-1}\sum_{j=1}^{n}(\DN_j-\avgDN)^2.
\label{eq_var}
\end{equation}
(The symbol $\sigma^{2}$ is used to denote the true or theoretical variance,
while $s^{2}$ is the estimated or ``sample'' variance.)

The variance estimate $\ssDN$ is subject to some uncertainty since only a limited number of independent spin configurations can be sampled.
Textbooks on statistics, \eg Ref. \cite{lehmann98}, show that the standard error of $\ssDN$ is
\begin{equation}
\stdssDN = \left(\frac{2}{n-1}\right)^{\frac{1}{2}} \varDN \approx \left(\frac{2}{n-1}\right)^{\frac{1}{2}} N.
\label{eq_varerror}
\end{equation}
If this ``spin noise'' is the only noise present, the overall signal-to-noise ratio (SNR) for the variance determination
depends only on $n$,
%
\begin{equation}
\SNR \equiv \frac{\varDN}{\stdssDN} = \left(\frac{n-1}{2}\right)^{\frac{1}{2}}.
\label{eq_snr_nonoise}
\end{equation}
This equation reveals a basic strategy for statistical spin detection: one should rapidly sample as many independent spin configurations as possible. Independent spin configurations can be obtained by periodically re-randomizing the ensemble.

The above analysis of the SNR represents an idealized case. In real experiments, the spin polarization is measured via an intermediate
quantity, such as the magnetic force, and the measurements are corrupted by noise. 
In our experiments, the measurement noise, including the cantilever thermal noise, is significant, and the spin signals have
a finite correlation time. These factors must be taken into account when analyzing the signal statistics. 

\begin{figure}[tb!]
      \begin{center}
      \includegraphics[width=0.50\textwidth]{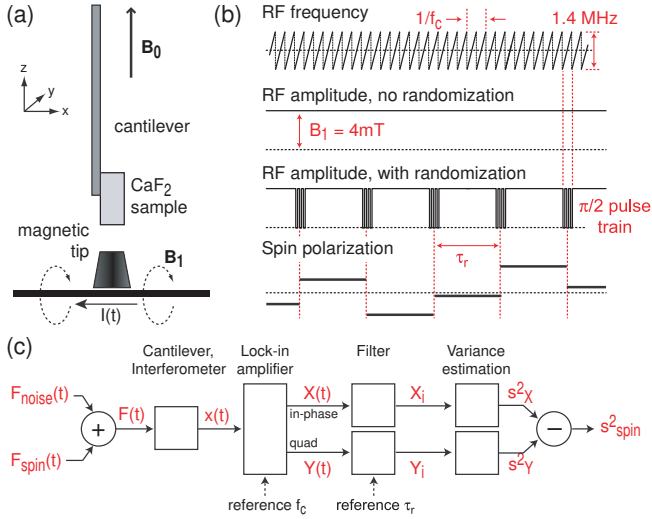}
      \end{center}
      \caption{
      (a) Experimental setup. A $(4\um)^3$ \CaF particle attached to the end of a single-crystal Si cantilever
      ($f_c=2.6\unit{kHz}$, $k_c=86\unit{\mu N/m}$, $Q=18,500$) is positioned $80\unit{nm}$ above a FeCo-magnetic tip
      and placed in a static external field $B_0=2.85\unit{T}$ ($B_0||\hat{z}$).
      \F spins are excited by an rf magnetic field $B_1=4\unit{mT}$ ($B_1||\hat{y}$),
      generated by passing current $I(t)$ through a microwire situated directly under the tip \cite{poggio07}.
      All experiments are carried out at $4.2\unit{K}$ and in high vacuum.
      (b) Protocol used for spin detection and randomization.
      Repetitive rf frequency sweeps with a center frequency of $114.8\unit{MHz}$
      invert the spins adiabatically twice per cantilever cycle \cite{slichter90a,madsen04}.
      For spin randomization, the rf field is interrupted at fixed intervals $\tr$ by a burst of $\pi/2$ pulses.
      Spin polarization is completely uncorrelated before and after randomization pulses.
      (c) Signal processing protocol as described in the text.
      }
      \label{fig_protocol}
\end{figure}

As shown in Fig. \ref{fig_protocol}, we observe the nuclear spin polarization using an ultra-sensitive cantilever to
detect the attonewton magnetic force between spins in the sample and a nearby magnetic tip \cite{sidles95,rugar94}.
The cantilever mechanical resonance is driven by cyclic adiabatic spin inversions induced by rf frequency sweeps,
and by thermal noise. A fiber-optic interferometer monitors
the resulting cantilever motion and the cantilever oscillation signal is synchronously detected by a two-channel
lock-in amplifier. The phase of the lock-in amplifier is set so that the spin signal plus thermal noise appear
in the in-phase (X) channel, while only thermal noise is present in the quadrature (Y) channel. The lock-in signals
are digitized and low-pass filtered in software so as to control the overall measurement bandwidth. 
Since the spin signal and the measurement noise are statistically independent, the variance of the in-phase channel is given by $\varX=\varspin+\varnoise$. The quadrature channel measures only noise, so
$\varY=\varnoise$. The spin portion of the variance is then given by $\varspin=\varX-\varY$.

In the experiment, the in-phase and quadrature channels are recorded for a time $\tottime $ and the spin signal variance $\ssspin$ is estimated from the measured sample variances according to $\ssspin=\ssX-\ssY$. 
We now seek an expression for the SNR associated with $\ssspin$ that is similar to Eq. (\ref{eq_snr_nonoise}), but also includes measurement noise. To obtain statistically independent spin configurations,
the spins are periodically randomized at intervals separated by time $\tr$ using bursts 
of short ($\pi /2$) rf pulses. The spin polarization remains essentially constant between randomizations if $\tr\ll\tm$, where $\tm$
is the correlation time of the naturally fluctuating spin polarization.
In order to keep the analysis simple
we assume that the X and Y channels are filtered using convolution-type filters that average the signals in the intervals between the periodic randomizations. The actual experiment used first order low pass filters instead of the assumed averaging filters, resulting in a slight reduction of SNR performance \cite{webb73}. 

The output of the averaging filters are assumed to be sampled synchronously with the randomizations, resulting in the measurement sequences $X_i$ and $Y_i$,
where $i=1,...,n$ and $n=\tottime /\tr$ is the number of independent spin configurations. 
From these samples, the variances $\ssX$ and $\ssY$
are calculated as in equation (\ref{eq_var}). The associated standard errors are
$\stdssX = \left[  2/ \left( n-1 \right) \right]^{1/2} \varX$ and 
$\stdssY = \left[ 2/ \left( n-1 \right) \right]^{1/2} \varY$.
The standard error of $\ssspin$ is then given by $\stdssspin=(\varssX+\varssY)^{1/2}$, or
$\stdssspin = \left\{ \frac{2}{n-1} [ (\varspin + \varnoise)^2 + (\varnoise)^2 ] \right\}^{1/2}$.
The SNR can then be written as
\begin{equation}
\SNR \equiv \frac{\varspin}{\stdssspin} = 
\left(\frac{\tottime / \tr - 1 }{2}\right)^{\frac{1}{2}} \left[1+2\frac{\varnoise}{\varspin}+2\frac{\varvarnoise}{\varvarspin}\right]^{-\frac{1}{2}}.
\label{eq_snr_withnoise}
\end{equation}
Note that in the absence of measurement noise, (\ref{eq_snr_withnoise}) is consistent with the previous result in (\ref{eq_snr_nonoise}).

To finish the analysis, we need to know how $\varnoise$ depends on the randomization interval $\tr$. For a convolution filter that averages the signal over the randomization interval $\tr$, the power transfer function is $\left| G(f)\right| ^2=\sin ^2 \left(\pi f \tr \right)/ \pi^2 f^2 \tr^2 $, which has an equivalent noise bandwidth of $\Delta f = 1/ 2 \tr$ (single-sided with units of Hz).
The noise variance  
is given by $\varnoise =  \Snoise \Delta f =  \Snoise / 2 \tr$, where $\Snoise$ is the single-sided power spectral density of the measurement noise in each channel.  We assume this spectral density to be independent of frequency within the bandwidth of the filter. This is the case in our experiment where the cantilever thermal noise has a relatively broad spectrum since the cantilever is strongly damped by feedback.
The SNR can thus be written as
\begin{equation}
\SNR = 
\left(\frac{\tottime / \tr - 1 }{2}\right)^{\frac{1}{2}} 
\left[1+\frac{\Snoise}{\tr \varspin}+\frac{\Snoise^{2}}{2 \tr^{2} \varvarspin}\right]^{-\frac{1}{2}}.
\label{eq_snr_withtauR}
\end{equation}
From the above equation, the SNR is found to be maximized when the randomization period is $\tr = \Snoise/\sqrt{2} \varspin$, which is equivalent to choosing
a randomization rate and associated filter bandwidth such that $\varnoise = \varspin /  \sqrt{2}$. With this choice of $\tr$ and assuming
that $\tottime \gg \tr$, we find
\begin{equation}
\SNR_\mr{max} =  \left(\frac{\varspin \tottime }{ 2 \left( \sqrt{2}+1 \right) \Snoise} \right)^{\frac{1}{2}}.
\label{eq_snr_max}
\end{equation}
\begin{figure}[tb!]
      \begin{center}
      \includegraphics[width=0.45\textwidth]{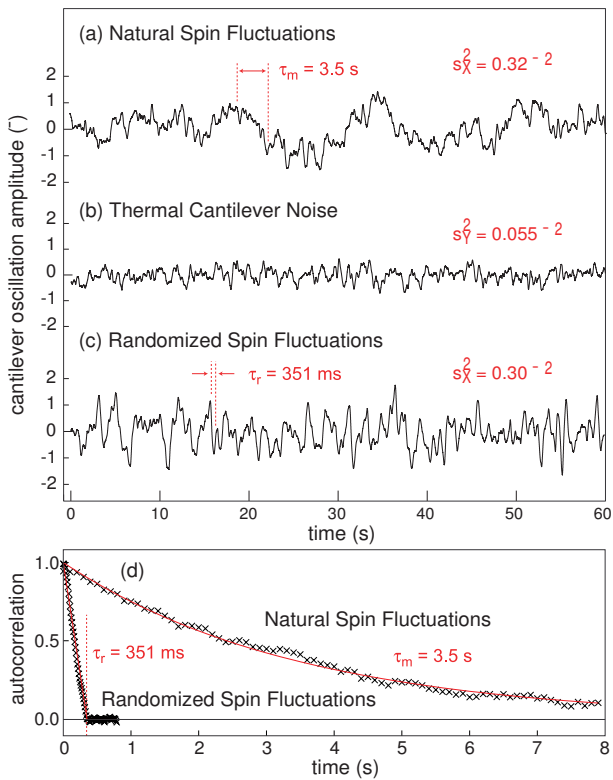}
      \end{center}
      \caption{
      Cantilever tip oscillation amplitude as a function of time.
      (a) In-phase signal $X(t)$ in the presence of the \F spins fluctuating naturally with a spin signal correlation time of $\tm\approx3.5\unit{s}$. 
      (b) Quadrature signal $Y(t)$ due to thermal fluctuations of the cantilever.
      The observed thermal noise correlation time ($\tau\approx200\unit{ms}$) is set by cantilever and filter bandwidths.
      (c) Same as (a), but spins are periodically re-randomized every $\tr=351\unit{ms}$.
      (d) Normalized autocorrelation function of spin signals obtained by analysis of an $8\unit{min}$ long record
      and after subtracting the measurement thermal noise autocorrelation function.
      }
      \label{fig_fluctuations}
\end{figure}

We have experimentally demonstrated the advantage of periodic spin randomization while measuring statistical polarization of \F in a \CaF single-crystal sample. In Fig. \ref{fig_fluctuations}(a) we show a typical 60 second duration record of $X(t)$ from the lock-in amplifier as the spins are cyclically inverted by rf frequency sweeps. This output contains contributions from both the spins and the thermal noise, while the quadrature channel signal $Y(t)$, shown in Fig. \ref{fig_fluctuations}(b), is predominantly just thermal noise. The X channel variance is $0.32\unit{\AA}^2$, which corresponds to a force variance of approximately $450\unit{aN}^2$. The Y channel variance is much smaller, about $0.055\unit{\AA}^2$, which corresponds to the thermal force variance of $78\unit{aN}^2$. Based on an estimated lateral field gradient of $10^6\unit{T/m}$, the observed spin fluctuations in the X channel correspond to an rms statistical polarization of about 1500 \F spins.
 
The spin fluctuations in Fig. \ref{fig_fluctuations}(a) can be seen by inspection to have long correlation times, on the order of seconds. By calculating the autocorrelation function associated with this and other similar time records, the autocorrelation was found to be well fit by an exponential decay with correlation time of $\tm \approx 3.5\unit{s}$ [Fig. \ref{fig_fluctuations}(d)].
$\tm$ is closely related to the intrinsic rotating-frame spin lifetime $T_{1\rho}$ and also depends on a number of extrinsic parameters,
for example the amplitude and modulation of the rf field $B_1$ \cite{slichter90,poggio07}.
If only a single one minute waveform, such as in Fig. \ref{fig_fluctuations}(a), is used to estimate the variance of $X(t)$, the error is large, approximately 35\%, since the number of independent samples $n= \tottime /\tm = 17$ is relatively small. 

The signal $X(t)$ obtained with periodic spin randomization is shown in Fig. \ref{fig_fluctuations}(c), where the
randomization is achieved by periodically interrupting the cyclic adiabatic frequency sweeps by a burst of $\pi/2$ pulses, as shown in Fig. \ref{fig_protocol}(c).
For this example we use the randomization period $\tr=351\unit{ms}$, which corresponds to a $\pi/2$-pulse burst
every $500\mr{th}$ adiabatic inversion. The burst consists of 20 pulses of $2\unit{\us}$ duration
with center frequencies uniformly distributed over the same frequency range as the adiabatic sweeps.
Judging from the autocorrelation function of the randomized fluctuations,
the spins are very effectively scrambled by this protocol [Fig. \ref{fig_fluctuations}(d)].
As expected, the autocorrelation function falls linearly and reaches zero at $\tr$.

The periodic randomization allows us to make an improved estimate of the spin signal variance.
Based on a randomization repetition period $\tr = 351\unit{ms}$ and the same total measurement time of $60\unit{s}$,
the number of independent measurements is now
$\tottime /\tr = 171$, which results in an uncertainty in the variance of about
$11\%$, an improvement of more than a factor of three.

\begin{figure}[tb!]
      \begin{center}
      \includegraphics[width=0.45\textwidth]{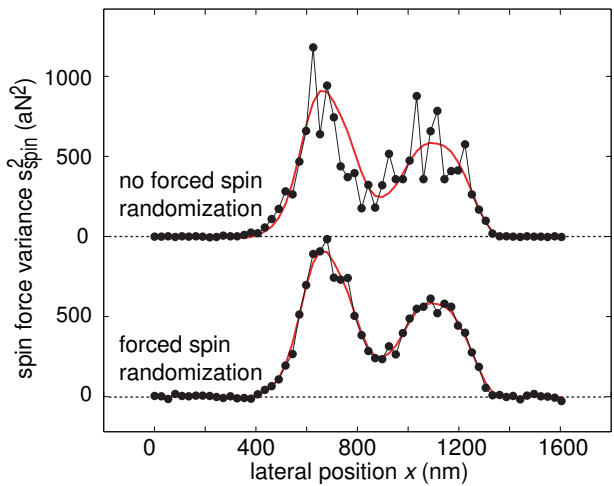}
      \end{center}
      \caption{One-dimensional image scans in $\hat{x}$-direction of the calcium fluoride sample.
      The points are the measured variances of the spin force and the solid lines are guides for the eye. Scatter of the data points in the top trace is due to the uncertainty of the variance estimates. Much improved SNR is seen in the lower trace where the spins are periodically randomized in order to rapidly probe a large number of independent spin configurations. Lock-in filter time constants for the top and bottom traces are $t_c=700\unit{ms}$ and $t_c=30\unit{ms}$, respectively.
      }
      \label{fig_1dscan}
\end{figure}
To more concretely illustrate the impact of periodic randomization on the SNR, we show in Fig. \ref{fig_1dscan}(a) a one-dimensional lateral imaging scan over the \CaF object.
Each point represents the estimated spin signal variance from a one minute data record.
Two overall signal maxima are visible that result from the two lobes of the imaging point spread function \cite{zuger96}. 
For the scan with natural spin randomization, the large uncertainty in the variance results in large scatter in the scan data over regions where the spin signal is highest. The periodic spin randomization, performed with $\tr = 113 \unit{ms}$, results in dramatic reduction of the data scatter, especially for the points with the largest spin signal [Fig. \ref{fig_1dscan}(b)]. The slightly larger noise in the regions where there is no spin signal is due to increased measurement noise that results from the larger measurement bandwidth needed to accommodate the shortened spin correlation time. 

\begin{figure}[tb!]
      \begin{center}
      \includegraphics[width=0.40\textwidth]{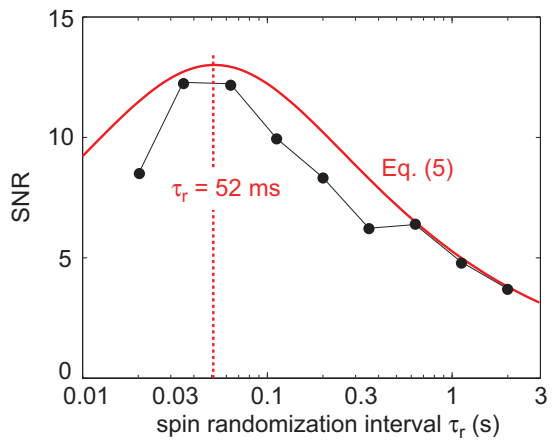}
      \end{center}
      \caption{Data points showing SNR for one minute measurements as a function of the randomization interval $\tr$.
      The solid line represents the expected SNR based on Eq. (\ref{eq_snr_withtauR}) using known experimental parameters (no free fit parameters).
      At the optimum $\tr=52\unit{ms}$, spin noise power equals $\sqrt{2}$-times the measurement noise power.
      }
      \label{fig_snr}
\end{figure}
As indicated by Eq. (\ref{eq_snr_withtauR}) and shown in Fig. \ref{fig_snr}, we can vary $\tr$ in order to optimize the SNR.
Here we measure the spin fluctuations at a position of large spin signal (the 670-nm position of the lateral scan shown in Fig. \ref{fig_1dscan}).
To determine each SNR data point, 50 one-minute data records were acquired. The sample variance $\ssspin$ from each record
was calculated and the SNR was determined by ${\rm{SNR}} = \overline{\ssspin} / s_{\ssspin}$,
where $\overline{\ssspin}$ is the mean of the sample variances and $s_{\ssspin}$ is the standard deviation of the variances.
As shorter randomization intervals are used, the SNR increases until it reaches a maximum around $\tr = 50\unit{ms}$.
The SNR then drops for the shortest $\tr$ value tested, as expected, since the large associated measurement bandwidth is admitting more measurement noise than is optimal. Obtaining results for 
even shorter $\tr$ was hampered by the finite response time of the cantilever ($12\unit{ms}$ for the cantilever damped by negative feedback). At the optimum point, the achieved SNR is 12, or about 6 times better than the SNR achieved without any forced spin randomization.

Figure \ref{fig_snr} also plots the theoretical SNR as predicted by Eq. (\ref{eq_snr_withtauR}) using the following parameters determined from the experiment:
$\varspin = 900\unit{aN^2}$ and $\Snoise = 66\unit{aN^2/Hz}$.
The theory predicts an optimal SNR of roughly 13 for a one-minute measurement time when $\tr = 52\unit{ms}$, in good agreement with the experimental results.
Overall, the SNR values obtained experimentally are just slightly lower than predicted, probably because the filtering that was performed did not use the ideal convolution filter assumed in the analysis, but rather a first order low pass filter of equal noise bandwidth (time constant $t_c=\tr/2$).

In conclusion, we have shown that noise inherent in the measurement of small, statistically polarized spin ensembles
can be mitigated by rapid re-randomization of the spins.
Beyond the fundamental interest in the control and measurement of spin ensembles in the nanometer regime,
the practical impact on nanoscale magnetic resonance imaging is significant:
the demonstrated $6 \times$ improvement in SNR allows a $36 \times$ increase of imaging speed.

We thank C. Rettner, M. Hart and M. Farinelli for fabrication of microwire and magnetic tip,
B. W. Chui for cantilever fabrication, and D. Pearson and
B. Melior for technical support. We acknowledge support from the DARPA QUIST
program administered through the US Army Research Office,
and the Stanford-IBM Center for Probing the Nanoscale, a NSF Nanoscale
Science and Engineering Center. C.L.D. acknowledges funding from
the Swiss National Science Foundation.

\end{document}